\documentclass{nature}

\usepackage{color}
\usepackage[libertine]{newtxmath}
\usepackage{caption}
\usepackage[misc]{ifsym}
\usepackage{graphicx}
\usepackage[colorlinks, linkcolor=blue, anchorcolor=blue, citecolor= blue]{hyperref}

\title{Robust photon blockade with hybrid molecular optomechanics}

\author{Jian Tang$^{1}$, Baijun Li$^{2}$, Bin Yin$^{1}$, Tian-Xiang Lu$^{3}$, Ran Huang$^{4~\text{\Letter}}$, Franco Nori$^{4,5}$ \& Hui Jing$^{1,6~\text{\Letter}}$}

\begin{document}

\maketitle

\begin{affiliations}
 \item Key Laboratory of Low-Dimensional Quantum Structures and Quantum Control of Ministry of Education, Hunan Normal University, Changsha 410081, China
 \item Research Center for Quantum Physics, Huzhou University, Huzhou 313000, China
 \item College of Physics and Electronic Information, Gannan Normal University, Ganzhou 341000, Jiangxi, China
 \item Quantum Information Physics Theory Research Team, Quantum Computing Center, RIKEN, Wako-shi, Saitama 351-0198, Japan
 \item Physics Department, The University of Michigan, Ann Arbor, Michigan 48109-1040, United States
 \item Institute for Quantum Science and Technology, College of Science, NUDT, Changsha 410073, P.R.China
 \item[$^\text{\Letter}$]{email}: ran.huang@riken.jp; jinghui73@foxmail.com
\end{affiliations}

\begin{abstract}
Molecular cavity optomechanical systems, featuring ultrahigh vibrational frequencies and strong light-matter interactions, hold significant promise for advancing applications in quantum science and technology. Specifically, by introducing metallic nanoparticles into microcavities, hybrid molecular cavity optomechanical systems can further enhance optical quality factors and system tunabilities, which enables scalable and controllable quantum platforms. In this study, we propose how to realize robust photon blockade, i.e., strong photon antibunching with arbitrary detuning conditions, by combining degenerate optical parametric amplification with a hybrid molecular cavity optomechanical system. More interesting, we find near-perfect optomechanical photon blockade at room temperature, which is robust against temperature and optical dissipation.  In addition, our approach can release the strict condition of high temporal resolution by combining features of conventional and unconventional photon blockade. Our approach offers a feasible route to study intriguing quantum effects in hybrid molecular cavity optomechanical systems, and holds promise for applications in nonclassical state engineering, quantum sensing, and photonic precision measurements.
\end{abstract}
		
\section*{\label{sec:level1}Introduction}
Cavity optomechanical systems~\cite{Aspelmeyer2014Cavity, Metcalfe2014Applications}, characterized by the interaction between optical fields and mechanical vibrations, offer a versatile platform for both fundamental research in physics and potential applications in high-precision sensing~\cite{Aspelmeyer2012Quantum, Kippenberg2008Cavity,Brennecke2008Cavity, Pirkkalainen2015Cavity,Li2021Cavity,Zhu2023Cavity, Xu2024Single}. Specifically, molecular vibrations, serving as nanoscale mechanical oscillators, exhibit exceptionally high resonant frequencies and strong optomechanical interactions with optical fields~\cite{Benz2016Single,Roelli2016MCT,Esteban2022Molecular}. These features enable molecular cavity optomechanical systems to remarkably resistant to temperature fluctuations and variations in system parameters. Therefore, molecular cavity optomechanical systems are ideal for applications such as heat transfer~\cite{Ashrafi2019Optomechanical}, optomechanically
induced transparency~\cite{Bin2025Molecular}, frequency up-conversion~\cite{Zou2024Amplifying,Chen2021Continuous,Xomalis2021Detecting}, metrology~\cite{Liu2018Room,Liu2018Coupled,Tabatabaei2015Tunable}, surface-enhanced Raman scattering~\cite{Zhang2020Optomechanical}, and molecular-protein analysis~\cite{Sadhanasatish2023Molecular}, as well as quantum heat engine~\cite{zhu2024autonomousquantumheatengine}, and entanglement~\cite{Huang2012Collective}.

Recent technological advances have enabled the integration of nanoparticles into high-quality microcavity platforms~\cite{Zhang2024Plasmonic}. Hybrid molecular cavity optomechanical systems~\cite{Shlesinger2023Hybrid,Shlesinger2021Integrated, Dezfouli2019Molecular} have been proposed and studied using metallic nanoparticles combined with Fabry-Pérot resonators~\cite{Shlesinger2023Hybrid}, microdisk cavities~\cite{Shlesinger2021Integrated}, or photonic crystals~\cite{Dezfouli2019Molecular}. These hybrid platforms not only retain the high resonant frequencies and strong optomechanical interactions, but also exhibit higher optical quality factors~\cite{Shlesinger2023Hybrid,Shlesinger2021Integrated,Dezfouli2019Molecular,Barreda2021Cavity}, exhibiting broad potential for various applications, such as microfilters and miniature spectrometers~\cite{zhang2024miniature}. More importantly, the platforms enhance the scalability of molecular optomechanical systems, enabling flexible control via coupled cavity configurations.

Optical parametric amplification is a second-order nonlinear process, in which an optical signal is amplified by a pump via the generation of an idler field~\cite{osti1984}, and can be used to realize sources of biphotons~\cite{uren2006generation,Mosley2008Heralded,Harris2007Chirp}, squeezed light~\cite{Crouch1988Broadband,Kashiwazaki2020Continuous,Qin2022Beating,kawasaki2024broadband}, and highly nonclassical states of light~\cite{YAN2021Generation,hou2011experimental}. Despite the inherently weak second-order nonlinearity, the large parametric gain can be achieved through strong pumping. Optical parametric amplification has been used to realize nonreciprocal optical transistor~\cite{Tang2022Quantum}, phonon lasing~\cite{lu2024quantum}, as well as entanglement between two atoms~\cite{Qin2018Exponentially}, long-lived or entangled cat states~\cite{Qin2021Generating,Chen2021Shortcuts}, tripartite quantum entanglement~\cite{Jiao2024Tripartite}, quantum sensing~\cite{Wang2024Quantum,Zhang2024Optical}, and photon blockade~\cite{Lv2015Squeezed,Wang2023Cavity,Shen2023Tunable}. In the weak pump regime, the nonlinear process can be seen as a different excitation path~\cite{Wang2023Cavity}. This nonlinear process gives rise to destructive quantum interference between transition pathways, which is the fundamental mechanism of unconventional photon blockade~\cite{Liew2010Single,Snijders2018Observation,Vaneph2018Observation,Flayac2017Unconventional,Jabri2022Enhanced,Bamba2011Origin,Li2019NUP}. Such nonlinear process induced photon blockade has been predicted in various platforms, including spinning cavity~\cite{Shen2020Nonreciprocal}, coupled cavities~\cite{Gou2023Simultaneous}, and cavity optomechanics~\cite{Wang2020oe,Xie2024Photon}.

In this work, we propose how to realize robust photon blockade scheme by introducing a degenerate optical parametric amplifier within a hybrid molecular cavity optomechanical system. Specifically, by tuning the intensity and phase of the parametric gain, we can achieve strong photon antibunching under arbitrary detuning conditions. More interesting, we find near-perfect optomechanical photon blockade at room temperature, which is robust against temperature and optical dissipation. Furthermore, we find the oscillations in $g^{(2)}(\tau)$ can be eliminated, thus release the strict condition of high temporal resolution in experimental detection. Our approach offers a feasible route to achieve robust photon blockade using existing hybrid molecular molecular cavity optomechanical platforms and holds promise for applications in nonclassical state engineering, quantum sensing, and photonic precision measurements.

\begin{figure*}[t!]
  \includegraphics[width=0.55\textwidth]{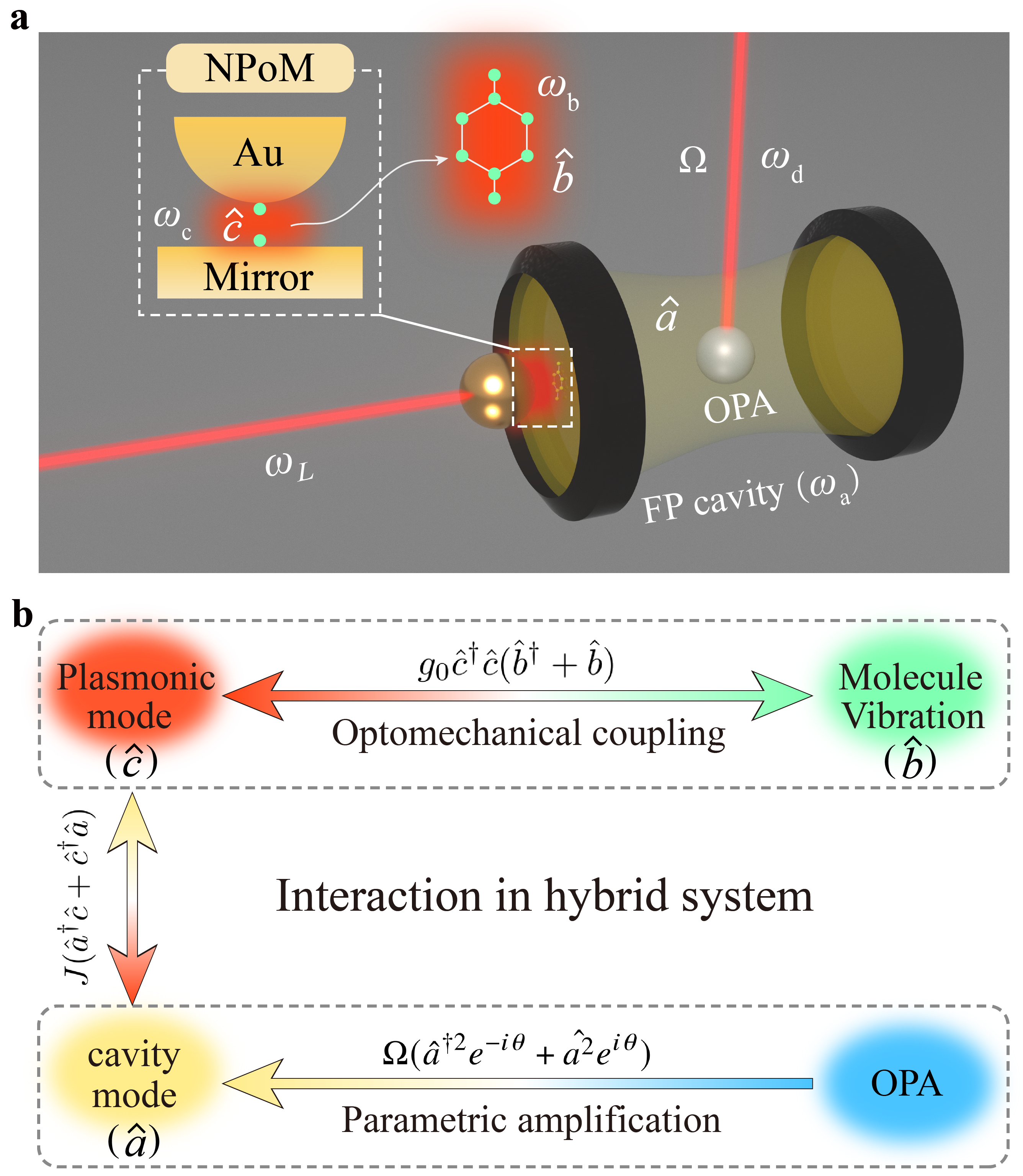}
  \centering
  \caption{{\textbf{Parametric amplification assistant hybrid molecule cavity optomechanical system.}}
  {\textbf{a}} Schematic of the hybrid molecule cavity optomechanical system. A plasmonic nanocavity, formed by metallic nanoparticles and molecules deposited on a Fabry-Pérot (FP) cavity, is driven by an external laser of frequency $\omega_{L}$. A degenerate optical parametric amplifier (OPA) inside the Fabry-Pérot cavity is driven by a laser of amplitude $\Omega$ and frequency $\omega_{d}$. The inset shows the nanoparticle-on-mirror (NPoM) structure of the plasmonic nanocavity, where molecules is situated in the nanogap. \textbf{b} Schematic diagram of the equivalent mode-coupling model. The plasmonic nanocavity mode is coupled to the molecular vibrational (acoustic) mode via radiation pressure. This mode is also coupled to the Fabry-Pérot cavity mode (incorporating the parametric process) through evanescent fields interactions.}
  \label{Fig1}
\end{figure*}
\section*{Model}
We consider the following hybrid molecular cavity optomechanical system. A nanoparticle-on-mirror (NPoM) structure consists of a metallic nanoparticle deposited on a mirror, which forms a plasmonic nanocavity that resonates at frequency $\omega_c$, as shown in the inset in Fig.~\ref{Fig1}a~\cite{Peng2024Construction,Hu2022Nanoparticle,Li2024Boosting}. A molecule is positioned in the nanoscale gap between the nanoparticle and the mirror, and is considered as a harmonic oscillator with a resonance frequency $\omega_b$ and a damping rate $\gamma$. This  molecule couples to the plasmonic field via optomechanical interaction. By introducing an additional mirror, a Fabry-Pérot cavity is formed with resonance frequency $\omega_c$, which contains a degenerate optical parametric amplifier. These components together form the hybrid molecular optomechanical system, as shown in Fig.~\ref{Fig1}a.

To achieve photon antibunching, we coherently drive the plasmonic nanocavity with a weak external field at frequency $\omega_L$. Through the coupling between the two cavities (with strength $J$), this excitation populates the Fabry-Pérot cavity mode, which then interacts with the optical parametric amplifier driven by an external laser at frequency $\omega_p$. The Hamiltonian for this three-mode system can be expressed as (assuming $\hbar = 1$):
\begin{eqnarray}
\hat{H}_1 &=& \omega_a \hat{a}^\dagger \hat{a} + \omega_c \hat{c}^\dagger \hat{c} + \omega_b \hat{b}^\dagger \hat{b} + \hat{H}_j+\hat{H}_d,  \nonumber\\
\hat{H}_j &=& J (\hat{a}^\dagger \hat{c} + \hat{c}^\dagger \hat{a})
+ g\hat{c}^{\dagger}\hat{c}(\hat{b}^{\dagger}+\hat{b}),  \nonumber\\
\hat{H}_d &=& \xi (\hat{c} e^{i\omega_L t}+\mathrm{H.c.}) +\Omega (\hat{a}^2 e^{i\theta} e^{i\omega_p t} +\mathrm{H.c.}),
\end{eqnarray}
where $\hat{a} (\hat{a}^\dagger)$, $\hat{c} (\hat{c}^\dagger)$, and $\hat{b} (\hat{b}^\dagger)$ represent the annihilation (creation) operators for the Fabry-Pérot cavity mode, the plasmonic nanocavity mode, and the molecular vibrational mode (phonons), respectively. Here, $g\hat{c}^{\dagger}\hat{c}(\hat{b}^{\dagger}+\hat{b})$ describes the optomechanical coupling interaction between the plasmonic nanocavity mode and the molecular vibrations. The coupling strength is given by $g = -\omega_c R_b \sqrt{\hbar / 2 m \omega_b} / (\varepsilon_0 V_c)$~\cite{Roelli2016MCT,Esteban2022Molecular}, where $R_b$ is the Raman polarizability, $V_c$ is the effective volume of the plasmonic nanocavity mode, and $\varepsilon_0$ is the vacuum permittivity. In this work, we take as $R_b = 3 \times 10^{-10}~ \varepsilon_0^2\text{\AA}^4\mathrm{amu}^{-1}$, and $V_c = 2.5~\mu\mathrm{m}^{-7}$~\cite{Roelli2016MCT,Esteban2022Molecular}, which yield $g/2\pi = 30~\mathrm{GHz}$. The optomechanical coupling coefficient $g$ is estimated on the order between from $2\pi\times10\ \mathrm{GHz}$ to $2\pi\times100\ \mathrm{GHz}$, and from $\omega_b/2\pi \sim6\ \mathrm{THz}$ to $48\ \mathrm{THz}$~\cite{Benz2016Single,Roelli2016MCT,Esteban2022Molecular}.
The driving amplitude $\xi$ is expressed as $\xi = \left[\gamma_{\mathrm{ex}} P_{\mathrm{in}} / (\hbar \omega_L)\right]^{1/2}$, where $P_{\mathrm{in}}$ is the input driving power, and $\gamma_{\mathrm{ex}}$ is the external loss rate. The other parameters in our work are: $\omega_a/2\pi=\omega_c/2\pi=460\ \mathrm{THz}$, $\omega_b/2\pi=30\ \mathrm{THz}$, $Q_a=10^{6}$, $Q_c=46$,
$\gamma/2\pi=30\ \mathrm{GHz}$, $J/2\pi=3\ \mathrm{THz}$. For the Fabry-Pérot cavities, the cavity quality factor $Q\sim10^{5}-10^{7}$~\cite{Aspelmeyer2014Cavity,Galinskiy2020Heterodyne}.

The final term in $\hat{H}_d$ represents the coupling between the cavity field and the parametric amplifier with the parametric gain amplitude $\Omega$ and phase $\theta$, which depends on the pump power and phase of the driving parametric amplifier, respectively~\cite{qin2018Exponentially}. To clearly see the interaction relation in this hybrid molecular optomechanical system, we further show the schematic diagram of the equivalent mode-coupling mode in Fig.~\ref{Fig1}b.

By transforming the system into a rotating frame at the driving frequency $\omega_L$ and assuming the condition $\omega_p = 2\omega_L$, the Hamiltonian of system reads
\begin{eqnarray}
\hat{H}_{2}= \Delta_{a}\hat{a}^{\dagger}\hat{a}+\Delta_{c}\hat{c}^{\dagger}\hat{c}+\omega_{m}\hat{b}^{\dagger}\hat{b}+\hat{H}_j+\xi(\hat{c}^{\dagger}+\hat{c})
+\Omega(e^{-i\text{\ensuremath{\theta}}}\hat{a}^{\dagger2}+e^{i\text{\ensuremath{\theta}}}\hat{a^{2}}),
\end{eqnarray}
where $\Delta_{a}=\omega_{a}-\omega_{l}$, $\Delta_{c}=\omega_{c}-\omega_{l}$. The quantum properties of light can be characterized by the second-order quantum correlation~\cite{Glauber1963Quantum}:
\begin{eqnarray}
g^{(2)}(\tau)=\lim_{t\to\infty}\frac{\langle\hat{a}_{1}^{\dagger}(t)\hat{a}_{1}^{\dagger}(t+\tau)\hat{a}_{1}(t+\tau)\hat{a}_{1}(t)\rangle}{\langle\hat{a}_{1}^{\dagger}(t)\hat{a}_{1}(t)\rangle^{2}},
\end{eqnarray}
which is usually measured by Hanbury Brown-Twiss interferometers~\cite{Birnbaum2005,Barak2008,Hamsen2018Strong,hennessy2007quantum,Faraon2008}. The condition $g^{(2)}(0)\ll1$ indicates single-photon blockade with sub-Poissonian photon-number statistics~\cite{Imamo1997Strongly,Birnbaum2005,scully1997quantum,agarwal2012quantum}. By introducing the density matrix of the cavity field $\rho(t)$ and Lindblad superoperator $D[O](\rho)=[O,\rho]=2O\rho O^{\dagger}-OO^{\dagger}\rho-\rho O^{\dagger}O$, (where $O=a,b,c$), this $g^{(2)}(\tau)$ [or $g^{(2)}(0)$] can be calculated by numerically solving the Lindblad master equation of this system~\cite{johansson2012qutip,johansson2013qutip2}:
\begin{eqnarray} \label{Eq:master}
L&=&	-i[H_{2},\rho]+\frac{\kappa_{a}(n_{{a}}+1)}{2}D[\hat{a}]+\frac{\kappa_{a}n_{{a}}}{2}D[\hat{a}^{\dagger}] \nonumber\\
&&+\frac{\kappa_{c}(n_{{c}}+1)}{2}D[\hat{c}]+\frac{\kappa_{c}n_{{c}}}{2}D[\hat{c}^{\dagger}] \nonumber\\
&&+\frac{\gamma(n_{{b}}+1)}{2}D[b]+\frac{\gamma n_{{b}}}{2}D[b^{\dagger}],
\end{eqnarray}
where $n_{{b(a,c)}}=[\exp(\hbar\omega_{{b(a,c)}}/(k_{b}{T}))-1)]^{-1}$ is the mean thermal phonon (photon) number of the mechanical (optical) mode at temperature $T$, with the Boltzmann constant $k_{b}$. The decay rate of the cavity (plasmonic) mode is given by $\kappa_{a}=\omega_{a}/Q_{a}$ ($\kappa_{c}=\omega_{c}/Q_{c}$).

\section*{Results}
To diagonalize the Hamiltonian, we introduce the following unitary operator $\hat{U}(\beta_{0}) = e^{\beta_{0} \hat{c}^\dagger \hat{c} (\hat{b}^\dagger - \hat{b})}$, where $\beta_{0} = -g_{0} / \omega_{m}$. Using the fact that $g_{0} \ll \omega_{m}$, we can safely neglect small exponential terms, allowing the mechanical mode of molecular vibrations to decouple from the plasmonic nanocavity mode. It is easy to make a unitary transformation on the Hamiltonian of the Molecular cavity optomechanical system:
\begin{eqnarray}
\hat{H}_{3}=\hat{U}\hat{H}_{2}\hat{U}^{\dagger}
&=&\Delta\hat{a}^{\dagger}\hat{a}+\Delta\hat{c}^{\dagger}\hat{c}+J\left(\hat{a}^{\dagger}\hat{c}+\hat{a}\hat{c}^{\dagger}\right)-G\left(\hat{c}^{\dagger}\hat{c}\right)^{2}\nonumber\\
&&+\xi\left(\hat{c}^{\dagger}+\hat{c}\right)+\Omega(\hat{a}^{\dagger2}e^{-i\theta}+\hat{a^{2}}e^{i\theta}),
\end{eqnarray}
where $G=\frac{g_{0}^{2}}{\omega_{m}}$. Here, we wrote in a frame where the pump and signal modes phase space rotate at frequency $\omega_{L}$. In accordance with the quantum trajectory method ~\cite{plenio1998quantumjump}, the decay of optical can be included in the effective Hamiltonian.
\begin{equation}
\hat{H}_{\text{eff}}=\hat{H}_{3}-\frac{i\kappa_{a}}{2}\hat{a}^{\dagger}\hat{a}-\frac{i\kappa_{c}}{2}\hat{c}^{\dagger}\hat{c}.
\end{equation}
Under the weak-driving condition($\xi\ll\gamma$), the Hilbert space can be restricted to $N=M+N=3$. The state of this system can be expressed as
\begin{eqnarray}
|\psi(t)\rangle=\sum_{N=0}^{3}\sum_{m=0}^{N}C_{m,N-m}|m,N-m\rangle,
\end{eqnarray}
with probability amplitudes $C_{m,N-m}$, which can be obtained by solving the Schrödinger equation
\begin{eqnarray}
i|\dot{\psi}(t)\rangle=\hat{H}_{\text{eff}}|\psi(t)\rangle.
\end{eqnarray}
Based on the effective Hamiltonian $\hat{H}_{\text{eff}}$ and the wave function, We can obtain the steady-state equations of the probability amplitudes. For ${C}_{00}(t)$, we have $i\dot{C}_{00}(t) =0$, while the other amplitudes fulfil the following equations:
\begin{eqnarray}
	0 &=&\Delta_{1}C_{1,0}+JC_{0,1}+\xi C_{1,1}, \nonumber\\
	0 &=&\Delta_{2}C_{0,1}+JC_{1,0}+\xi C_{0,0}+\sqrt{2}\xi C_{0,2},\nonumber\\
	0 &=&2\Delta_{1}C_{2,0}+\sqrt{2}JC_{1,1}+\sqrt{2}\Omega e^{i\text{\ensuremath{\theta}}}C_{0,0},\nonumber\\
	0 &=&(\Delta_{1}+\Delta_{2})C_{1,1}+\sqrt{2}JC_{0,2}+\sqrt{2}JC_{2,0}+\xi C_{1,0}, \nonumber\\
	0 &=&2\Delta_{2}C_{0,2}+\sqrt{2}JC_{1,1}+\sqrt{2}\xi C_{0,1}.
\end{eqnarray}
Here, we have $\Delta_{1}=\Delta_{c}-\frac{i\kappa_{a}}{2}$, $\Delta_{2}=\Delta_{c}-G-\frac{i\kappa_{c}}{2}$.

\begin{figure*}[t!]
\includegraphics[width=0.99\textwidth]{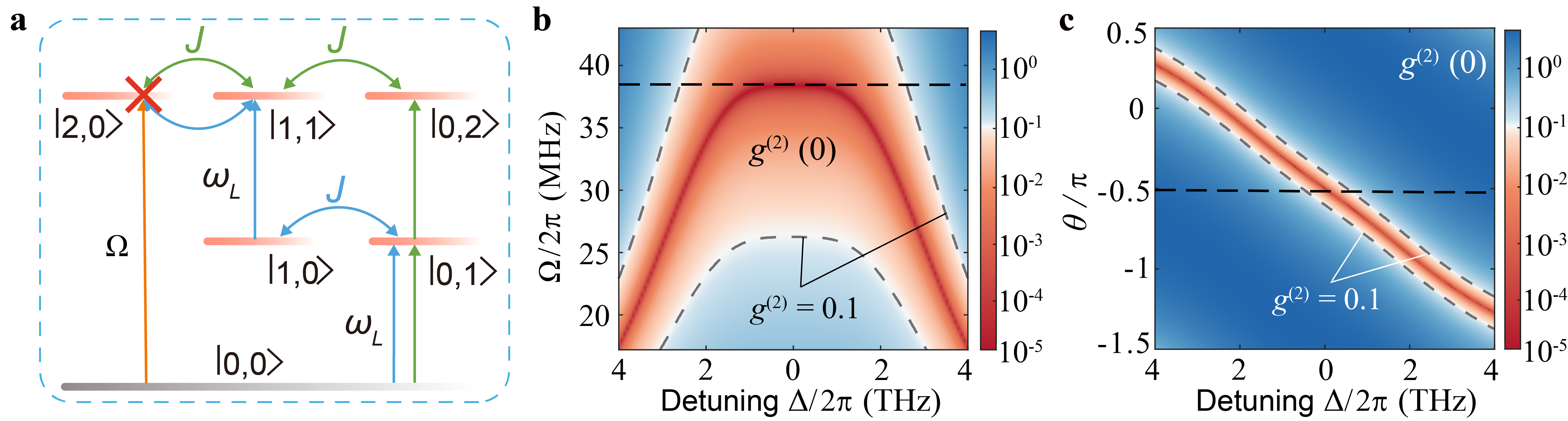}
  \caption{\textbf{The mechanism and optimal parametric gain conditions of molecular optomechanical photon blockade}.
  {\textbf{a}} {The destructive interference of the three transition from state $|0,0\rangle$ to state $|2,0\rangle$ (red, blue, green arrows) prevents two-photon occupation, enabling the unconventional photon blockade effect.}
  {\textbf{b}} {Second-order correlation $g^{(2)}(0)$ as a function versus detuning $\Delta$ and parametric gain amplitude $\Omega$.}
  {\textbf{c}} {Quantum correlation $g^{(2)}(0)$ as a function versus of detuning $\Delta$ and parametric gain phase $\theta$. Here, $P_\mathrm{in} = 0.1\,\mathrm{nW}$, $T=0 \mathrm{K}$. The other parameters are given in the main text.}
  }
  \label{Fig2}
\end{figure*}

We obtain the following solutions by considering the initial condition $C_{00}(0)=1$:
\begin{eqnarray}
C_{10} &=&\frac{-\xi \Delta}{\eta_1 }, \nonumber\\
C_{20} &=&\frac{\sqrt{2}\xi^{2}J^{2}(\Delta_{1}+\Delta_{2}-G)-\sqrt{2}\Omega e^{i\text{\ensuremath{\theta}}}\eta_{1}\eta_{2}}{\eta_{1}\eta_{3}},
\end{eqnarray}%
where $\eta_{1}=\Delta_{1}\Delta_{2}-J^{2}$, $\eta_{2}=\eta_{1}+\Delta_{2}^{2}-\text{(}\Delta_{1}+\Delta_{2})G$, $\eta_{3}=2\Delta_{1}\eta_{2}-2J^{2}(\Delta_{2}-G)$. Since $\left|C_{10}\right|^{2}\gg\left|C_{20}\right|^{2}$ and $\left|C_{10}\right|^{2}\gg\left|C_{11}\right|^{2}$, the second-order correlation function can be written as
\begin{eqnarray}
g^{(2)}(0)=\frac{2\left|C_{20}\right|^{2}}{\left(\left|C_{10}\right|^{2}+2\left|C_{20}\right|^{2}+\left|C_{11}\right|^{2}\right)^{2}}\approx\frac{2\left|C_{20}\right|^{2}}{\left|C_{10}\right|^{4}}.
\end{eqnarray}
The photon blockade can occur with $P_{20} = 0$~\cite{Liew2010Single,Bamba2011Origin,Flayac2017Unconventional}, which can be attributed to the destructive interference among three transition paths, as illustrated in Fig.~\ref{Fig2}a: $|0,0\rangle \stackrel{\Omega}{\longrightarrow} |2,0\rangle$, $|0,0\rangle \stackrel{\omega_{L}}{\longrightarrow} |0,1\rangle \stackrel{\omega_{L}}{\longrightarrow} |0,2\rangle \stackrel{\sqrt{2}J}{\longrightarrow} |1,1\rangle \stackrel{\sqrt{2}J}{\longrightarrow} |2,0\rangle$, and $|1,0\rangle \stackrel{J}{\longrightarrow} |0,1\rangle \stackrel{\omega_{L}}{\longrightarrow} |1,1\rangle \stackrel{\sqrt{2}J}{\longrightarrow} |2,0\rangle$. Thus, the optimal parametric gain satisfies
\begin{eqnarray}
\Omega_{\mathrm{opt}}
&=& \frac{\big|\xi^2 J^2 \left( \Delta_1 + \Delta_2 - G \right)\big|}{\big|\eta_1 \, \eta_2\big|}, \nonumber\\
\theta_{\mathrm{opt}}
&=& \arg\!\left[ \xi^2 J^2 \left( \Delta_1 + \Delta_2 - G \right) \right]
- \arg\!\left( \eta_1 \, \eta_2 \right).
\end{eqnarray}

We plot the second-order correlation function $g^{(2)}(0)$ as a function of detuning and parametric gain amplitude (or phase), as shown in Figs.~\ref{Fig2}b, c, where the curves represent $g^{(2)}(0)=0.1$. The robust photon blockade can be achieved across a wide range of detunings by appropriately tuning the parametric gain amplitude and phase. As the detuning $\Delta/2\pi$ decreases, the gain amplitude $\Omega$ must be increased to maintain the optimal blockade conditions.

\begin{figure}[t!]
	\centering
	\includegraphics[width=0.81\columnwidth]{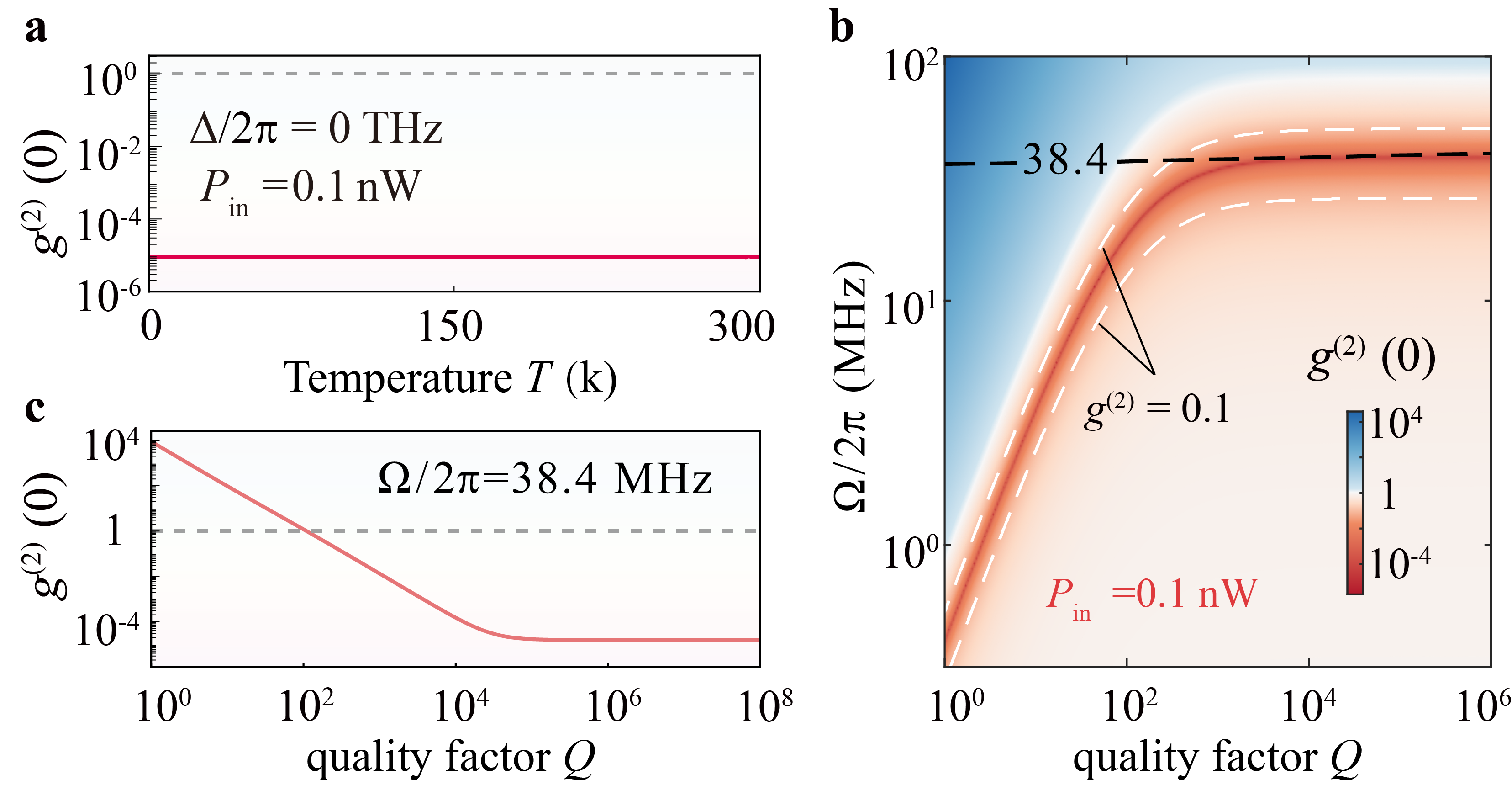}
	\caption{\textbf{Robustness of the molecular optomechanical photon blockade against varying temperature and cavity quality factor.}
\textbf{a} The second-order correlation function $g^{(2)}(0)$ versus $T$ at driving power $P_\mathrm{in} = 0.1\,\mathrm{nW}$ and $\Delta/2\pi = 0\,\mathrm{THz}$, showing that $g^{(2)}(0) \ll 1$ under room-temperature conditions, thereby demonstrating robust photon blockade in the presence of thermal effects.
\textbf{b} Influence of the cavity quality factor $Q$ on unconventional photon blockade at an input power of $P_\mathrm{in} = 0.1\,\mathrm{nW}$. Across a broad range of $Q$ values, perfect unconventional photon blockade can be achieved by properly tuning the $\Omega$, and the optimal $\Omega$ is consistent in the high-$Q$ regime.
\textbf{c} For $\Omega/2\pi = 38.4\,\mathrm{MHz}$, $g^{(2)}(0)$ remains effectively insensitive to $Q$ for $Q > 10^{5}$, demonstrating the minimal impact of the cavity quality factor on unconventional photon blockade under these conditions. The other parameters are the same as those in Fig.~\ref{Fig2}.}
	\label{Fig33}
\end{figure}

Our results show that photon blockade in this system exhibits remarkable robustness to temperature variations. As temperature increases, the thermal occupations of phonons ($n_{{b}}=[\exp(\hbar\omega_{{b}}/(k_{b}{T}))-1)]^{-1}$) rise, typically weakening photon blockade in conventional optomechanical systems. In contrast, the molecular optomechanical system features an acoustic mode frequency of molecular vibrations that is sufficiently high ($\omega_b/2\pi=30\ \mathrm{THz}$) to keep phonon occupation comparatively low ($n_{b} \approx 0.8$), even at room temperature. While the photon thermal occupation remains negligible, allowing the photon blockade effect to persist essentially undisturbed, as illustrated in Fig.~\ref{Fig33} a.

We further find that single-photon blockade is robust against the quality factor of the coupled Fabry-Pérot cavity. As illustrated in Fig.~\ref{Fig33}b, strong antibunching ($g_{2} \ll 1$) is observed at $Q = 1$ by tuning the parametric strength, indicating that single-photon blockade can be achieved even in low-$Q$ cavities. For high-$Q$ cavities, the optimal parametric strength for single-photon blockade stabilizes and does not vary significantly. Hence, as shown in Fig.~\ref{Fig33}c, once the Fabry-Pérot cavity quality factor $Q$ exceeds $10^{5}$, the parametric
gain for unconventional photon blockade remain nearly unchanged, eliminating the need for stringent cavity design or measurement to achieve robust single-photon blockade.

\begin{figure*}[t!]
	\centering
	\includegraphics[width=0.9\textwidth]{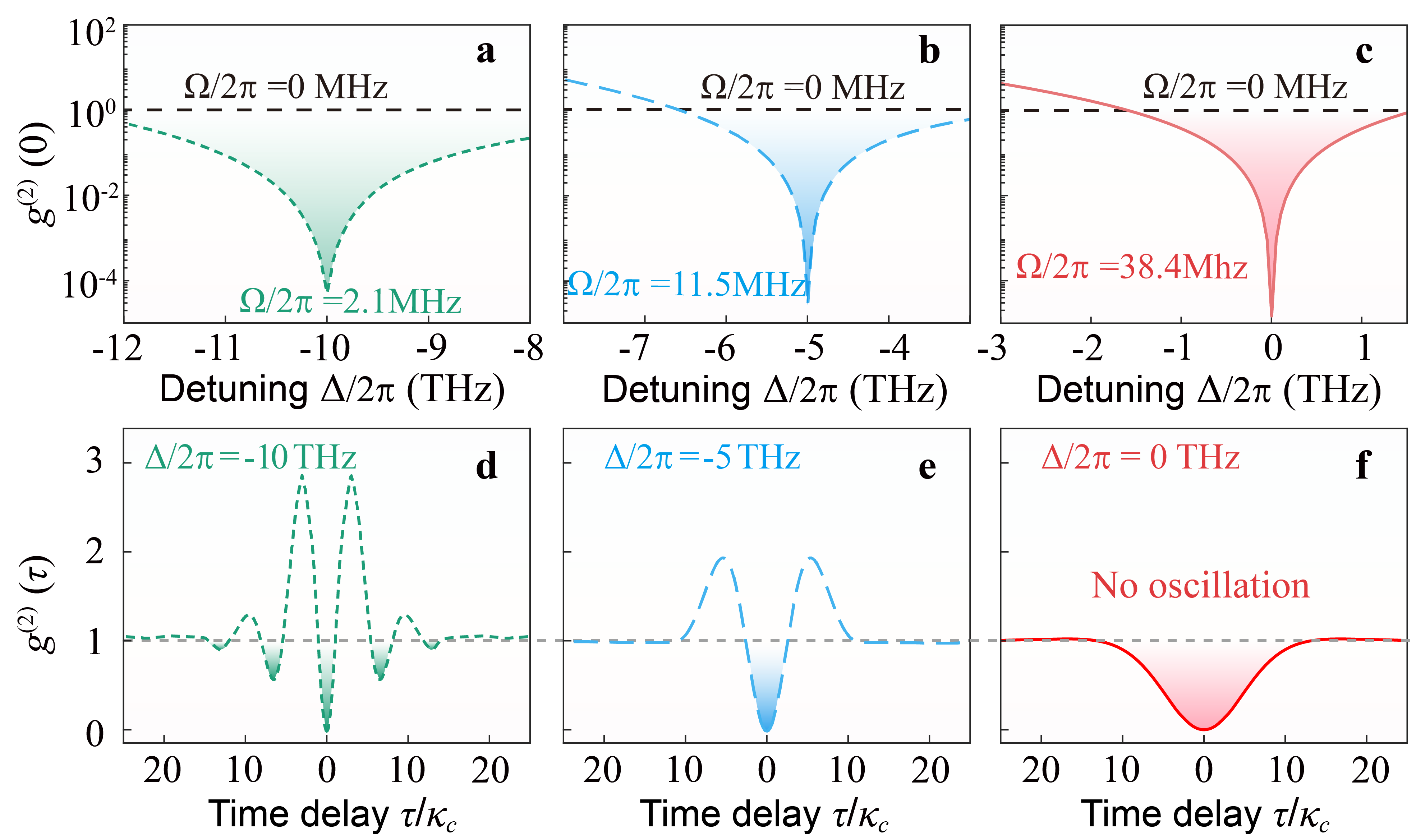}
	\caption{\textbf{Robust and oscillation-free photon blockade over time.} For \textbf{a} $\Delta/2\pi = -10\,\mathrm{THz}$, \textbf{b} $\Delta/2\pi = -5\,\mathrm{THz}$ and \textbf{c} $\Delta/2\pi = 0\,\mathrm{THz}$, the optimal parametric gain amplitude for realizing robust photon blockade is $\Omega/2\pi = 2.1\,\mathrm{MHz}$, $\Omega/2\pi = 11.5\,\mathrm{MHz}$ and $\Omega/2\pi = 38.4\,\mathrm{MHz}$, respectively.
\textbf{d, e, f} The evolution of the second-order correlation function $g^{(2)}(\tau)$ with respect to the time delay $\tau/\gamma$, showing no oscillations at $\Delta = 0$, thereby indicating that high temporal resolution is not needed to observe photon blockade in this regime. Here, $P_\mathrm{in} = 0.1\,\mathrm{nW}$, and the other parameters are the same as those in Fig.~\ref{Fig2}.
}
	\label{Fig4}
\end{figure*}

Finally, we find that our approach can release the strict condition of high temporal resolution in photon correlation measurements.  As the detuning \(\Delta\) decreases, the optimal parametric gain amplitude \(\Omega\) required to achieve photon blockade increases accordingly, while the time-dependent second-order correlation function \(g^{(2)}(\tau)\) gradually becomes oscillation-free. Specifically, for \(\Delta/2\pi = -10~\mathrm{THz}\) and \(-5~\mathrm{THz}\), robust photon blockade is obtained at \(\Omega/2\pi = 2.1~\mathrm{MHz}\) and \(11.5~\mathrm{MHz}\), respectively [Figs.~\ref{Fig4}a, b]. In these cases, although \(g^{(2)}(0)\) is significantly suppressed, oscillations still appear in \(g^{(2)}(\tau)\) during evolution [Figs.~\ref{Fig4}d, e]. Remarkably, at zero detuning (\(\Delta/2\pi = 0~\mathrm{THz}\)), the minimum of \(g^{(2)}(0)\) is achieved at \(\Omega/2\pi = 38.4~\mathrm{MHz}\) [Figs.~\ref{Fig4}c]. And the system exhibits features of both conventional and unconventional photon blockade: \(g^{(2)}(0)\) remains well below unity while the temporal oscillations in \(g^{(2)}(\tau)\) are completely eliminated [Figs.~\ref{Fig4}f].
As a result, the antibunching persists over a prolonged duration of nearly $10 \kappa_c$ ($\kappa_c / 2\pi = 10~\mathrm{THz}$), indicating that photon blockade can be observed reliably and without the need for high temporal resolution in detection.

\section*{DISCUSSION}

\noindent
In summary, we have proposed how to realize a robust photon blockade scheme based on a hybrid molecular cavity optomechanical system introducing a degenerate optical parametric amplifier. By tuning the parametric gain amplitude and phase, strong photon antibunching is achieved across a wide range of detuning. Owing to the high-frequency molecular vibrations and enhanced optomechanical coupling, the system exhibits remarkable resilience against variations in temperature and optical dissipation, maintaining nonclassical photon statistics even under room-temperature conditions. Furthermore, the system exhibits both conventional and unconventional photon blockade characteristics, i.e., \(g^{(2)}(0)\) remains well below unity while the temporal oscillations in \(g^{(2)}(\tau)\) are completely eliminated. This allows photon blockade to be observed without the need for high temporal resolution, improving experimental feasibility.

This work highlights the unique advantages of hybrid molecular cavity optomechanical systems, which combine nanoscale mechanical degrees of freedom with high optical quality factors and strong light–matter interactions. These features make such systems promising platforms for realizing a range of quantum effects beyond photon blockade, including light-motion entanglement~\cite{wang2014reservoir,Jiao2020Nonreciprocal,liu2024phase,liying2021vector}, optomechanical cat states~\cite{hauer2023nonlinear,li2023optomechanical,liao2016macroscopic}, photon bundles~\cite{munoz2014emitters,bin2020nphonon,bin2021parity,zou2023dynamical}, and thermally robust squeezed or correlated states~\cite{Qin2022Beating,kawasaki2024broadband}. Moreover, the intrinsic resilience of system to environmental decoherence makes it particularly attractive for applications in quantum sensing and precision measurement under ambient conditions. Our findings offer a foundation for exploring broader functionalities of hybrid optomechanical architectures in integrated quantum photonics.

\section*{Data Availability}
All relevant data that support the figures within this paper and other findings of this study are available from the corresponding authors upon reasonable request.

\section*{Code Availability}

All relevant code support the figures within this paper and other findings of this study are available from the corresponding authors upon reasonable request.

\begin{addendum}
    \item[Funding] H.J. is supported by the National Key R\&D Program (Grant No. 2024YFE0102400), the National Natural Science Foundation of China (Grants No.~11935006 and No.~12421005), the Hunan Major Sci-Tech Program (Grant No. 2023ZJ1010). R.H. is supported by the RIKEN Special Postdoctoral Researchers (SPDR) program. X.-W.X. is supported by the Innovation Program for Quantum Science and Technology (Grant No. 2024ZD0301000). T.-X.L. is supported by the NSFC (Grant No. 12565001, 12205054) and the Natural Science Foundation of Jiangxi Province (No. 20252BAC200163). F.N. is supported by the Japan Science and Technology Agency (JST) [via the CREST Quantum Frontiers program Grant No. JPMJCR24I2, the Quantum Leap Flagship Program (Q-LEAP), and the Moonshot R\&D Grant No. JPMJMS20611], the Office of Naval Research (ONR) Global (via Grant No. N62909-23-1-2074), and the National Science Foundation (NSF) NQVL: QSDT Award No. 2435166.
    \item[Acknowledgements] The authors thank Xun-Wei Xu for helpful discussions.
    \item[Author Contributions] H.J. designed and supervised the project. R.H. advised on the interpretation of the results. J.T. performed the research and wrote the manuscript. B.-J.L discussed the results. B.Y. prepared figure 1. F.N. and T.-X.L revised the manuscript. All authors discussed the results and reviewed the manuscript.
    \item[Competing Interests] The authors declare that they have no competing financial interests.
    \item[Correspondence] and requests for materials should be addressed to H. J and R.H.
\end{addendum}

\end{document}